\documentclass[journal=jacsat,manuscript=article]{achemso}

\usepackage[version=3]{mhchem} % Formula subscripts using \ce{}

\author{Yijue Ding}
\affiliation{Department of Physics, Kansas State University, Manhattan, KS 66502, USA}
\email{yijueding@gmail.com}

\title[An \textsf{achemso} demo]
  {Ultrafast photodissociation dynamics of dichloromethane on
three-dimensional potential energy surfaces and
its Coulomb explosion signature}

\abbreviations{IR,NMR,UV}
\keywords{American Chemical Society, \LaTeX}

\begin{document}


%\begin{tocentry}

%Some journals require a graphical entry for the Table of Contents.
%This should be laid out ``print ready'' so that the sizing of the

%Inside the \texttt{tocentry} environment, the font used is Helvetica
%8\,pt, as required by \emph{Journal of the American Chemical

%The surrounding frame is 9\,cm by 3.5\,cm, which is the maximum
%permitted for  \emph{Journal of the American Chemical Society}
%graphical table of content entries. The box will not resize if the
%content is too big: instead it will overflow the edge of the box.

%This box and the associated title will always be printed on a
%separate page at the end of the document.

%\end{tocentry}

\begin{abstract}
We present efficient and reliable molecular dynamics simulations of the photodissociation of dichloromethane followed by Coulomb explosion. These simulations are performed by calculating trajectories on accurate potential energy surfaces of the low-lying excited states of the neutral dichloromethane molecule. The subsequent time-resolved Coulomb explosions are simulated on the triply charged ionic state, assuming Coulomb interactions between ionic fragments. Both the neutral state trajectories and the simulated Coulomb explosion observables indicate that intra-molecular photoisomerization of dichloromethane is unlikely to occur. Estimating the kinetic energy release using \textit{ab initio} ionic potential reveals a discrepancy of approximately 5-8 eV compared to our simulated values using Coulomb potential. The molecular structural changes during photodissociation are clearly mapped to the ionic-fragment coincidence signals, demonstrating the Coulomb explosion imaging technique as a powerful tool to probe the time-resolved reaction dynamics. 
\end{abstract}

\section{Introduction}

Dichloromethane ($\text{CH}_2\text{Cl}_2$) is an intriguing organic molecule that has been studied due to its role in photochemical reactions and its relevance to environmental and industrial applications\cite{simmonds2006,hossaini2017}. Photochemical reactions of dichloromethane are typically triggered by vacuum ultraviolet (VUV) excitations. The corresponding absorption spectra have been measured over a broad range of 6 eV to 12 eV\cite{mandal2014,lange2020}. Following VUV absorption, the excited molecule undergoes photodissociation, leading to the cleavage of C-Cl bonds and the formation of reactive fragments such as $\text{CH}_2$ and Cl radicals.

Recent advances in ultrafast imaging techniques, such as Coulomb explosion imaging (CEI)\cite{cei1,cei2} and ultrafast electron diffraction (UED)\cite{ued}, have opened new avenues for exploring the real-time dynamics of molecular dissociation. The CEI technique relies on detecting the spatial distribution of charged fragments produced by intense laser ionization. It has been employed to identify distinct photochemical reaction channels\cite{ziaee2023,ding2024,surgendu2024}, isolate peculiar roaming processes\cite{endo2020}, and investigate the Jahn-Teller effect\cite{li2021,wang2024}.

The photodissociation of excited molecules involves a complex interplay of electronic, nuclear, and ionization dynamics, which are often studied to investigate non-adiabatic transitions, bond cleavage mechanisms, and quantum coherence\cite{carlos2004,polli2010,ding2016,ding2017}. To the best of our knowledge, no molecular dynamics simulations have been performed for the photodissociation of $\text{CH}_2\text{Cl}_2$. Apart from absorption spectra, \textit{ab initio} calculations on this molecule are very limited. Only a few potential energy curves have been calculated\cite{xiao2007,almasy2019}, and a few key geometries of the ground state have been identified\cite{lewars1998}.
Although some photodissociation experiments have been conducted\cite{tonokura1992,brownsword1997,hayakawa2008}, no ultrafast laser experiments have directly explored the time-resolved dynamics of $\text{CH}_2\text{Cl}_2$. Specifically, the Coulomb explosion experiment has been conducted to unveil the static structure of $\text{CH}_2\text{Cl}_2$\cite{gagnon2008}, but no pump-probe experiments using CEI as the probe have been reported so far.

In this work, we perform molecular dynamics simulation for the VUV-induced photodissociation of $\text{CH}_2\text{Cl}_2$ probed by three-fragment CEI, as the corresponding time-resolved CEI experiment in our lab is on the schedule\cite{privatecomm}.
%aiming to inspire the corresponding ultrafast laser experiments.
We simulate the photodissociation by calculating trajectories on three-dimensional \textit{ab initio} potential energy surfaces (PESs) of the neutral molecule. The subsequent Coulomb explosion at different pump-probe delays is simulated on the triply charged ionic state, assuming pure Coulomb interactions between ionic fragments.

\section{Computational methods}
\subsection{Dichloromethane electronic structures}
The electronic structure of $\text{CH}_2\text{Cl}_2$ is calculated using the coupled-cluster method with singles and doubles (CCSD)\cite{hampel1992}. The singlet excited-state energies are determined using the equation-of-motion (EOM) approach\cite{eomccsd}. In this work, the ground state and the four lowest excited singlet states are calculated using the EOM-CCSD method, and the calculations are performed using the MOLPRO quantum chemistry package\cite{molpro,molpro2}. The Dunning-type cc-pVTZ basis sets\cite{dunningbasis} are employed for all atoms in the \textit{ab initio} calculations.
Unlike other halogen species, the spin-orbit coupling (SOC) in chlorine atoms is relatively small; the energy difference between the Cl$(^2\text{P}_{3/2})$ and Cl$(^2\text{P}_{1/2})$ states is only 0.1 eV\cite{uehara1987}. Therefore, SOC effects are neglected in our calculations.

Geometry optimization is performed following the ground-state CCSD calculation to obtain the equilibrium geometry of the $\text{CH}_2\text{Cl}_2$ ground state. The resulting geometric parameters are $R_{\text{CH}}=1.08\text{\AA}$, $\Theta_{\text{HCH}}=111.5^\circ$, $R_{\text{CCl}}=1.77\text{\AA}$, and $\Theta_{\text{ClCCl}}=112^\circ$.
At this equilibrium geometry, excitation energies and corresponding oscillator strengths are calculated using the EOM-CCSD method and compared with values from the literature, as shown in Table \ref{table1}. The excitation energies are slightly higher than those obtained using multi-reference methods but remain within the experimental spectrum, as the valence excitation spectrum is broad.
\begin{table}
  \caption{Calculated excitation energies (in eV) and corresponding oscillator strengths (in parentheses) for the four lowest singlet excited states of $\text{CH}_2\text{Cl}_2$ in $C_s$ symmetry, compared with theoretical and experimental results from the literature. }
  \label{table1}
  \begin{tabular}{lllll}
    \hline
    State  & This work\textsuperscript{\emph{a}} &  Theo.\textsuperscript{\emph{b}} & Theo.\textsuperscript{\emph{c}} & Exp.\textsuperscript{\emph{d}} \\
    \hline
    1$A''$ & 7.18(0.0011) & 7.18(0.0002)& 6.71(0.007)&  7.07 \\
    1$A'$  & 7.26(0.0217)  & 7.31(0.0117)& 6.80(0.022)&  7.52 \\
    2$A''$ & 7.53(0.0000) & 7.60(0.0000) & &   \\
    2$A'$ & 7.73(0.0053)  & 7.81(0.0046)& &  7.96\\
    \hline
  \end{tabular}

  \textsuperscript{\emph{a}} EOM-CCSD/cc-pVTZ
  \textsuperscript{\emph{b}} EOM-CCSD/aug-cc-pV5Z(g)+R, Ref. \citenum{lange2020}
  \textsuperscript{\emph{c}} MS-CASPT2(12,10)/ANORCC, Ref. \citenum{xiao2007}
  \textsuperscript{\emph{d}} Ref. \citenum{lange2020}
\end{table}

\begin{figure}
  \includegraphics[width=0.6\textwidth]{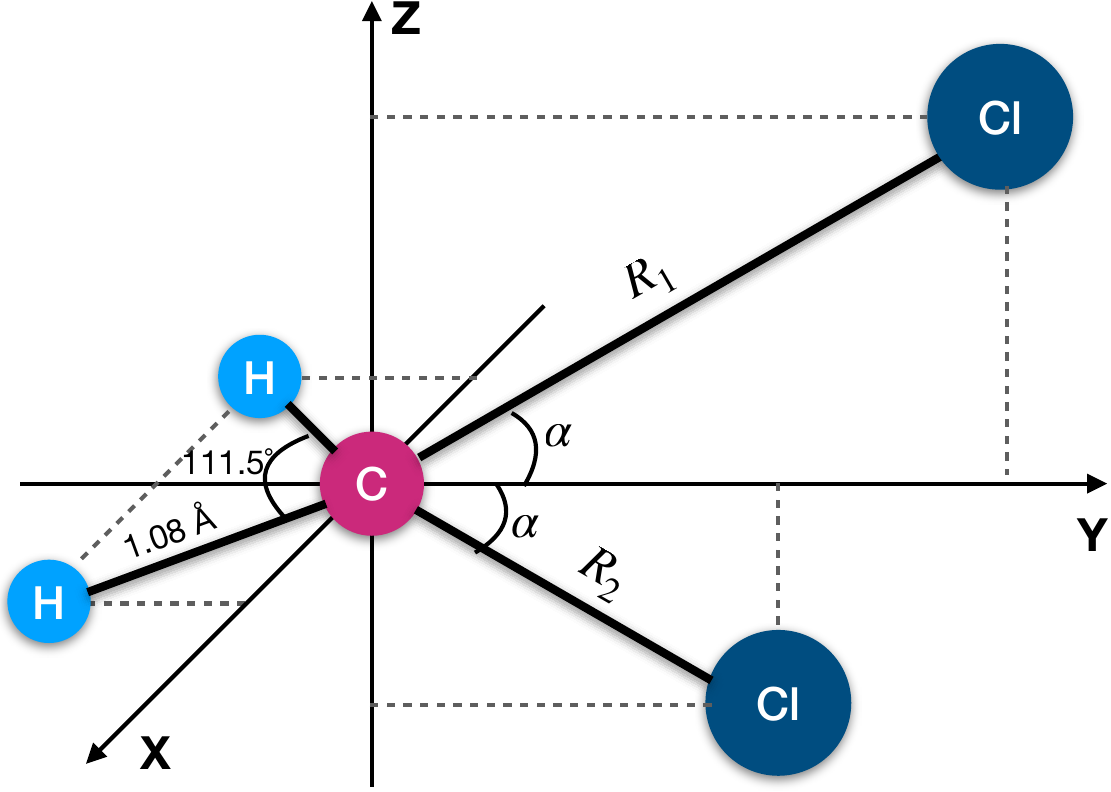}
  \caption{Geometric parameters of $\text{CH}_2\text{Cl}_2$ shown in its body frame. The two H atoms are constrained to lie in the x-y plane while the two Cl atoms are constrained to lie in the y-z plane. The two C-Cl bond lengths $R_1$ and $R_2$, and half of the ClCCl angle $\alpha$ constitute the three degrees of freedom used for building the potential energy surfaces.}
  \label{fig1}
\end{figure}

Geometric constraints are imposed on the molecule to reduce the dimensionality of the PES. As shown in Fig. \ref{fig1}, these constraints are as follows: 1) The C–H bond lengths and the H–C–H bond angle are fixed at their equilibrium geometry values. 2) The $\text{CH}_2$ plane is constrained to remain perpendicular to the $\text{CCl}_2$ plane. 3) The $\text{CH}_2$ plane is required to bisect the Cl–C–Cl bond angle. 4) The $\text{CCl}_2$ plane is required to bisect the H–C–H bond angle.
With these constraints, the molecule is reduced to three internal degrees of freedom: the two C–Cl bond lengths, $R_1$ and $R_2$, and half of the Cl–C–Cl bond angle, $\alpha$. These constraints also preserve the $C_s$ symmetry of the molecule across the entire PES.
By simplifying the PES and the nuclear dynamics, these constraints reduce computational complexity without significantly affecting the main physical processes during photodissociation, such as bond breaking and bond rearrangements between the carbon and chlorine atoms.

\subsection{Interpolating the potential energy surfaces}
The potential energy surfaces are interpolated using a tri-cubic spline method\cite{ding2024,tricubicspline}. It is equivalent to sequential interpolations on each dimension, but is implemented in a streamlined approach. 

At first, the ab initio energy points are calculated on a structured grid that can be represented by
\begin{equation}
\begin{split}
    \{R_1^{(1)},\dots,R_1^{(i)},R_1^{(i+1)},\dots,R_1^{(M)}\}\otimes\\
    \{R_2^{(1)},\dots,R_2^{(j)},R_2^{(j+1)},\dots,R_2^{(N)}\}\otimes\\
    \{\alpha^{(1)},\dots,\alpha^{(k)},\alpha^{(k+1)},\dots,\alpha^{(L)}\},
\end{split}
\label{eq3}
\end{equation}
where $M,N,L$ are the number of grid points in each dimension. The entire volume where the potential function is defined is sliced into small cubes by this structured grid. Within each cube, the potential function is defined as
\begin{equation}
    V(\Tilde{R}_1,\Tilde{R}_2,\Tilde{\alpha})= \sum_{m,n,l=0}^3 c_{mnl}\Tilde{R}_1^m\Tilde{R}_2^n\Tilde{\alpha}^l\quad(0\le \Tilde{R}_1<1,0\le \Tilde{R}_2<1,0\le \Tilde{\alpha}<1),
    \label{eqinterp}
\end{equation}
where the rescaled variables are
\begin{equation}
    \Tilde{R}_1=\frac{R_1-R_1^{(i)}}{R_1^{(i+1)}-R_1^{(i)}},\Tilde{R}_2=\frac{R_2-R_2^{(j)}}{R_2^{(j+1)}-R_2^{(j)}},\Tilde{\alpha}=\frac{\alpha-\alpha^{(k)}}{\alpha^{(k+1)}-\alpha^{(k)}}.
\end{equation}
On the other hand, the following potential energy (obtained from \textit{ab initio} calculation) and its derivatives (obtained from 1D interpolation) are evaluated at the 8 grid points 
 $(\Tilde{R}_1,\Tilde{R}_2,\Tilde{\alpha})=(0,0,0),(0,0,1),\cdots,(1,1,1)$: 
%$(\{R_1^{(i)},R_2^{(j)},\alpha^{(k)}\},\dots , \{R_1^{(i+1)},R_2^{(j+1)},\alpha^{(k+1)}\})$:
\begin{equation}
    \{V,\frac{\partial V}{\partial \Tilde{R}_1},\frac{\partial V}{\partial \Tilde{R}_2},\frac{\partial V}{\partial \Tilde{\alpha}},\frac{\partial^2 V}{\partial \Tilde{R}_1\partial \Tilde{R}_2},\frac{\partial^2 V}{\partial \Tilde{R}_1\partial \Tilde{\alpha}},\frac{\partial^2 V}{\partial \Tilde{R}_2\partial \Tilde{\alpha}},\frac{\partial^3 V}{\partial \Tilde{R}_1\partial \Tilde{R}_2 \partial \Tilde{\alpha}}\}.
    \label{eq4}
\end{equation}
The coefficients $c_{mnl}$, represented by a vector $\textbf{c}$, and the values in Eq. \eqref{eq4}, represented by a vector $\textbf{b}$, are connected by a universal constant matrix $\textbf{B}$:
\begin{equation}
    \textbf{Bc}=\textbf{b}
\end{equation}
The interpolated potential and its first-order derivatives are continuous throughout the volume where the potential function is defined. Figure 2 show the potential curves and the corresponding derivatives along the 3 coordinates. 
\begin{figure}
  \includegraphics[width=\textwidth]{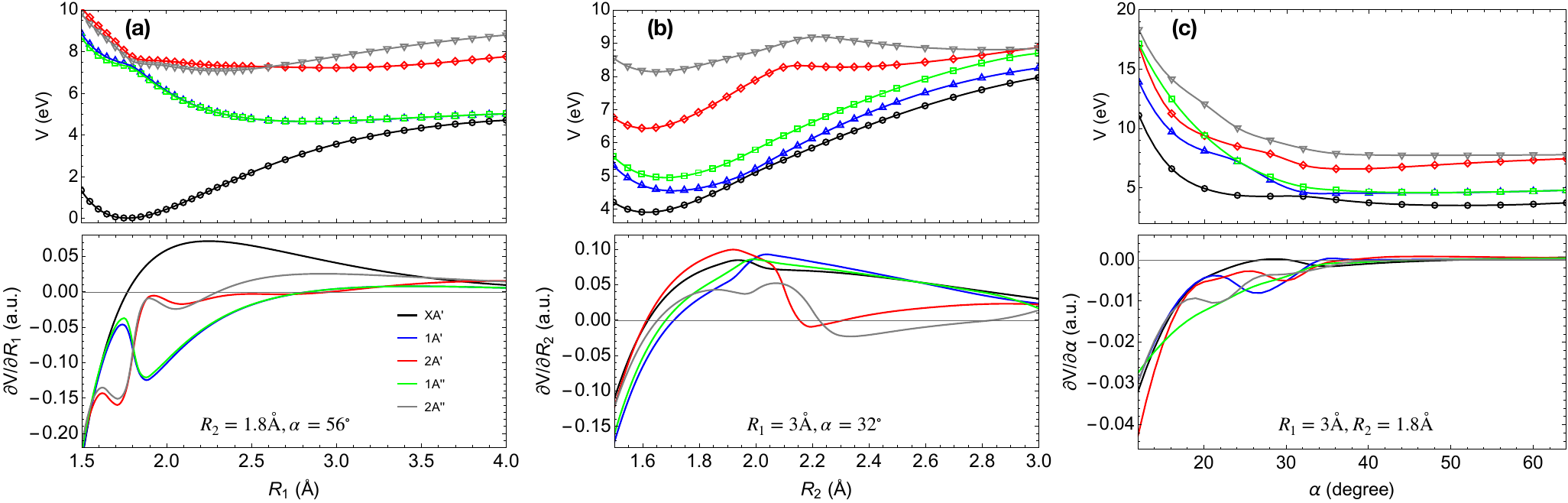}
  \caption{Potential energies and the corresponding derivatives for the lowest 5 electronic states of $\text{CH}_2\text{Cl}_2$ in $C_s$ symmetry as a function of $R_1$ (a), $R_2$ (b), and $\alpha$ (c) with the other degrees of freedom fixed at the values specified in each panel. The open markers denote the original energy data from \textit{ab initio} calculations, while the solid curves denote the tri-cubic spline interpolation.}
  \label{fig2}
\end{figure}

%The potential energy is calculated up to $R_1=4\text{ \AA}$ ($R_1$ is treated as the dissociation coordinate). Thus, the potential is extrapolated to characterize the long range behavior. To extrapolate the potential, the potential is written as follows, 
The long-range potential energy is extrapolated using a van der Waals form, which is written as
%\begin{equation}
 % V^{LR}_{\text{CH}_2\text{Cl}_2}(R_1,R_2,\alpha) = V_{\text{CH}_2\text{Cl}}(R_2,\alpha) - %\frac{C^{(1)}_6}{R_1^6} - \frac{C^{(2)}_6}{R_\text{ClCl}^6}+V_{th},
 % \label{eqlr}
%\end{equation}

\begin{equation}
  V^{LR}(R_1,R_2,\alpha) = V_{th}(R_2,\alpha) -\frac{C_6(R_2,\alpha)}{R_1^6},
  \label{eqlr}
\end{equation}
where the coefficients $C_6(R_2,\alpha)$ and $V_{th}(R_2,\alpha)$ are obtained by fitting the long-range tails of the \textit{ab initio} potential energy points and are interpolated over $R_2$ and $\alpha$ using the cubic spline interpolation method.
%for $R_1>3.5 \text{ \AA}$ and $R_\text{ClCl}>3.5 \text{ \AA}$.

%The entire PES is obtained by merging the interpolated potential function Eq.~\eqref{eqinterp} and the long range potential function Eq.~\eqref{eqlr} using a function
%\begin{equation}
 %   V=\frac{\exp[-(R-4a_0)/a_0]}{1+\exp[-(R-4a_0)/a_0]}V^{SR}+\frac{1}{1+\exp[-(R-4a_0)/a_0]}V^{LR}
%\end{equation}
%where the switch function is selected to be 
%\begin{equation}
 %   s=\frac{1}{1+\exp[-(R-4a_0)/a_0]}
%\end{equation}

\subsection{Classical trajectory calculations}

We simulate nuclear dynamics classically by solving the Euler-Lagrange equation of motion, which is given by
\begin{equation}
    \frac{d}{dt} \frac{\partial L}{\partial \dot{q}_i}-\frac{\partial L}{\partial q_i}=0 \quad (i=1\cdots 6), 
    \label{eqeom}
\end{equation}
where the 6 degrees of freedom (DOF) are $\{q_1\cdots q_6\}=\{\psi,\theta,\phi, R_1, R_2, \alpha\}$, and $\dot{q}_i=dq_i/dt$ denotes the generalized velocity. Apart from the 3 internal DOF $R_1$, $R_2$ and $\alpha$, we also include the three Euler angles $\psi$, $\theta$ and $\phi$ to describe the overall rotation of the entire molecule. 

The initial conditions are generated from a thermal Wigner distribution of the normal coordinates of the ground state near the equilibrium at an effective temperature T=300 K. The three normal coordinates $\eta_i(i=1,2,3)$ can be written as $\{\eta_1,\eta_2,\eta_3\}=\textbf{U}\{R_1,R_2,\alpha\}$, where $\textbf{U}$ is the transformation matrix. The thermal Wigner probability distribution is then given by
\begin{equation}
    W(\eta,\dot{\eta})=\frac{1}{(\pi\hbar)^3}\prod_{i=1}^3\exp\left[-\frac{\eta_i^2}{2\sigma_i^2}-\frac{2\sigma_i^2\dot{\eta_i}^2}{\hbar^2}\right],
    \label{eqwigner}
\end{equation}
where $\sigma_i(i=1,2,3)$ is a parameter that determines the width of the distribution for each normal coordinate. It is written as 
\begin{equation}
    \sigma_i=\sqrt{\frac{\hbar}{2\omega_i}\coth\left(\frac{\hbar\omega_i}{2k_BT}\right)},
\end{equation}
where $k_B$ is the Boltzmann constant and $\omega_i$ is the normal mode frequency. Our normal mode analysis on the ground-state PES yields $\{\omega_1,\omega_2,\omega_3\}=\{287,738,811\}\text{ cm}^{-1}$, corresponding to Cl–C–Cl bending, C–Cl symmetric stretching, and C–Cl asymmetric stretching modes, respectively. These frequencies are in overall agreement with previously reported experimental and theoretical values, as shown in Table \ref{table2}. The only noticeable discrepancy is the asymmetric stretching mode, which is approximately 6\% higher than the experimental value. Additionally, the molecule is prepared in its rotational ground state with random orientation.

\begin{table}
  \caption{Harmonic vibrational frequencies (cm$^{-1}$) of $\text{CH}_2\text{Cl}_2$ obtained from the interpolated ground-state PES compared with previously reported theoretical and experimental values.}
  \label{table2}
  \begin{tabular}{lllll}
    \hline
    Normal mode & This work \textsuperscript{\emph{a}}& Theo.\textsuperscript{\emph{b}}  & Theo.\textsuperscript{\emph{c}}  & Exp.\textsuperscript{\emph{d}}\\
    \hline
    Cl-C-Cl bend 	&	 287	&	286 & 290 & 282 \\
    C-$\text{Cl}_2$ sym. str. 	&	 738	&	734 & 713  & 713 \\
    C-$\text{Cl}_2$ asym. str.	&	 811	&	773 & 757&  760 \\
    \hline
  \end{tabular}
  
  \textsuperscript{\emph{a}} From interpolated PES.
  \textsuperscript{\emph{b}} PBE0/aug-cc-pV5Z, Ref. \citenum{mandal2014};
  \textsuperscript{\emph{c}} CASSCF/ANO-RCC, Ref. \citenum{xiao2007}.
  \textsuperscript{\emph{d}} Ref. \citenum{duncan1986}.
\end{table}

In a Franck-Condon (FC) transition, the molecule transitions to valence excited states by absorbing a single VUV photon, with the momentum and position of each nucleus remaining unchanged. Thus, we propagate trajectories on excited-state PESs with initial conditions sampled from the ground-state PES.
The lowest four excited states (1-2$A'$ and 1-2$A''$) are energetically accessible via absorption of a 158 nm photon near the FC region, as shown in Table \ref{table1}. However, molecules in the 2$A'$ and 2$A''$ states are non-dissociative, and the 2$A''$ state exhibits zero oscillator strength with the ground state. Consequently, dissociation from these states can only occur through non-adiabatic transitions to lower states near the $R_1 = R_2$ conical intersection, where the molecule adopts $C_{2v}$ symmetry. Such a mechanism has been studied in $\text{CH}_2\text{I}_2$, where Liu \textit{et al.} referred to it as indirect dissociation\cite{liu2020}.
In this work, we focus on direct dissociation and simulate molecular dynamics exclusively for the dissociative 1$A'$ and 1$A''$ states. We calculate 1000 trajectories up to 150 fs for each adiabatic state by integrating the Euler-Lagrange equation Eq. \eqref{eqeom} using an adaptive Runge-Kutta algorithm with a relative error tolerance of $10^{-7}$. Non-adiabatic transitions are not considered in our simulations because we observe no trajectories approach any conical intersections beyond the FC region. 

\section{Results and discussion}
\subsection{Nuclear dynamics on the excited state PESs}

One possible reaction channel for dihalomethane photodissociation is the so-called intra-molecular isomerization. Previous studies have proposed the minimal energy through conical intersection (MECI) mechanism to explain the seeming but doubtful isomerizations in $\text{CH}_2\text{I}_2$ and $\text{CH}\text{Br}_3$\cite{borin2016}. However, neither experiments nor molecular dynamics simulations have confirmed such a reaction channel or the proposed mechanism.

First, we discuss whether intra-molecular isomerization can occur in $\text{CH}_2\text{Cl}_2$ based on our trajectory calculations. The isomer represents a local minimum on the ground-state PES and is connected to the equilibrium geometry (Franck-Condon point) via a transition state (TS). The geometric parameters and corresponding potential energies of these states are listed in Table \ref{table3}.
The isomerization involves a C–Cl bond breaking process followed by Cl–Cl bond formation. 
We categorize two scenarios of isomerization: 1) The molecule transforms adiabatically on the excited states, but within a short period during photodissociation, the transient nuclear geometry resembles that of the isomer. 2) The molecule undergoes non-adiabatic transition to the ground state and becomes temporarily trapped in the isomer potential well before dissociation. The first scenario may not be strictly called isomerization but probing methods like UED and CEI may not distinguish between these two scenarios because the electronic states are not identified. 

\begin{table}
  \caption{Geometric parameters and corresponding energies of the Franck-Condon point, the transition state and the isomer state of $\text{CH}_2\text{Cl}_2$ on the ground electronic state. }
  \label{table3}
  \begin{tabular}{llllll}
    \hline
    State & $R_1$ (\AA) & $R_2$ (\AA) & $\alpha$ (deg) & $R_{ClCl}$ (\AA)& E (eV)   \\
    \hline
    FC   & 1.77 & 1.77 & 56 & 2.93 & 0  \\
    TS & 2.91 & 1.60 & 32  & 2.63 & 3.90 \\
    Isomer  & 3.49 & 1.60 & 18 & 2.39 & 3.41  \\
    \hline
  \end{tabular}
\end{table}

\begin{figure}
  \includegraphics[width=0.6\textwidth]{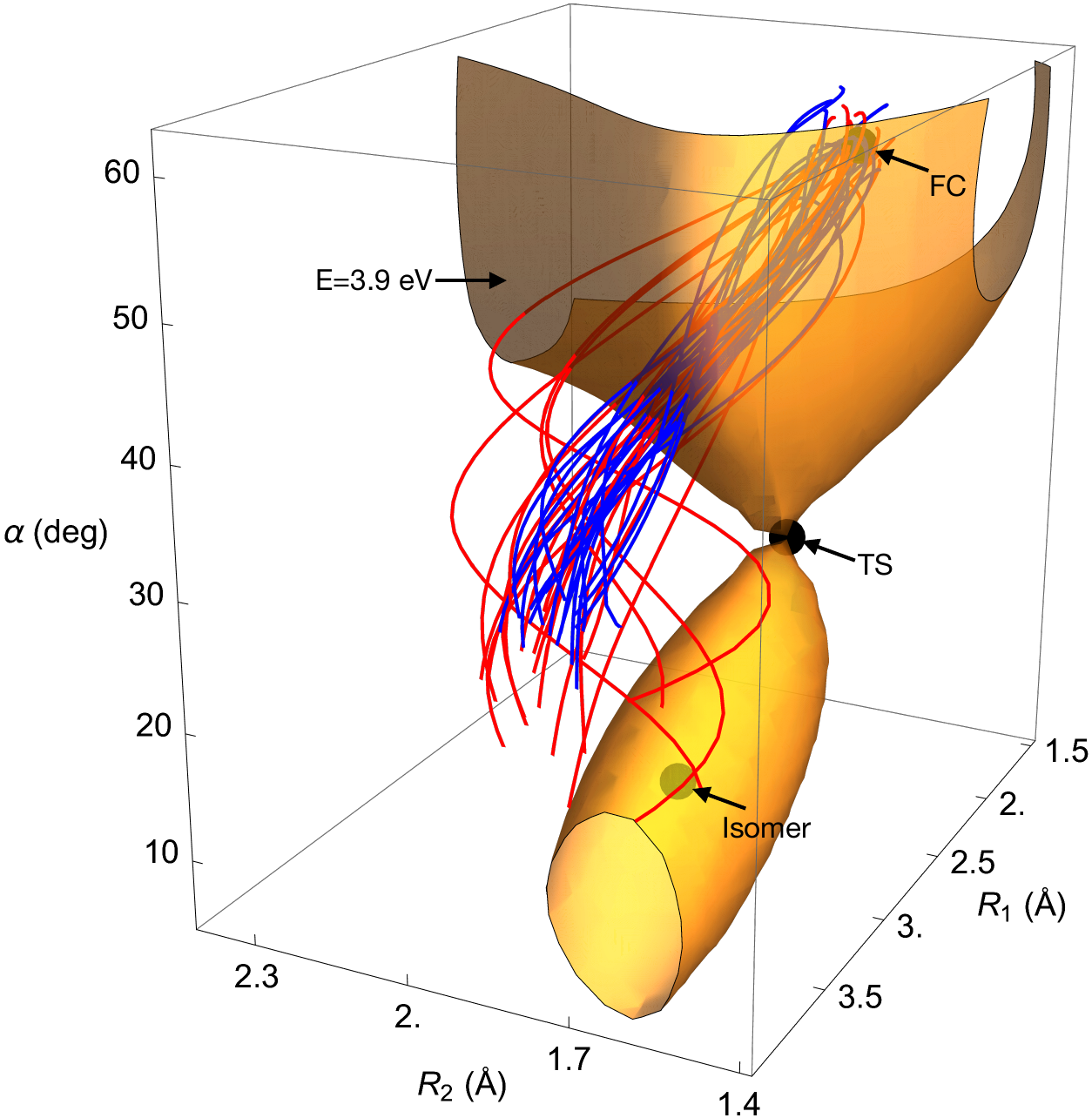}
  \caption{Visualization of 40 representative trajectories calculated on the $1A'$ (red) and $1A''$ (blue) states (20 for each state), and the iso-energy contour plot of the ground PES at $V(R_1,R_2,\alpha)=3.9$ eV. The geometries of the Franck-Condon point, the transition state and the isomer are also shown (black dots). The dissociation coordinate is $R_1$. Note that $R_1$ and $R_2$ coordinates are shown on different scales. }
  \label{fig3}
\end{figure}

Our adiabatic trajectory calculations on the 1$A'$ and 1$A''$ states reveal that isomerization in the first scenario is unlikely to occur. We observe no trajectories forming $\text{CH}_2\text{Cl}$-Cl isomer-like geometries during photodissociation, as illustrated by 40 representative trajectories in Fig. \ref{fig3}.
To further investigate, we define a geometric region with boundaries at the TS energy $V(R_1, R_2, \alpha) = 3.9$ eV, as shown by the iso-energy surface in Fig. \ref{fig3}. The volume enclosing the isomer, with energy 3.41 eV$<V<$3.9 eV, includes not only all possible vibrational states of the isomer but also two-body dissociative states. Even under this alleviated condition, we still find no trajectories entering this volume.
Regarding the second scenario, we observe no trajectories on the 1$A'$ state approaching any conical intersections with the ground state, indicating that non-adiabatic transitions to the ground state are unlikely to occur.

\begin{figure}
  \includegraphics[width=\textwidth]{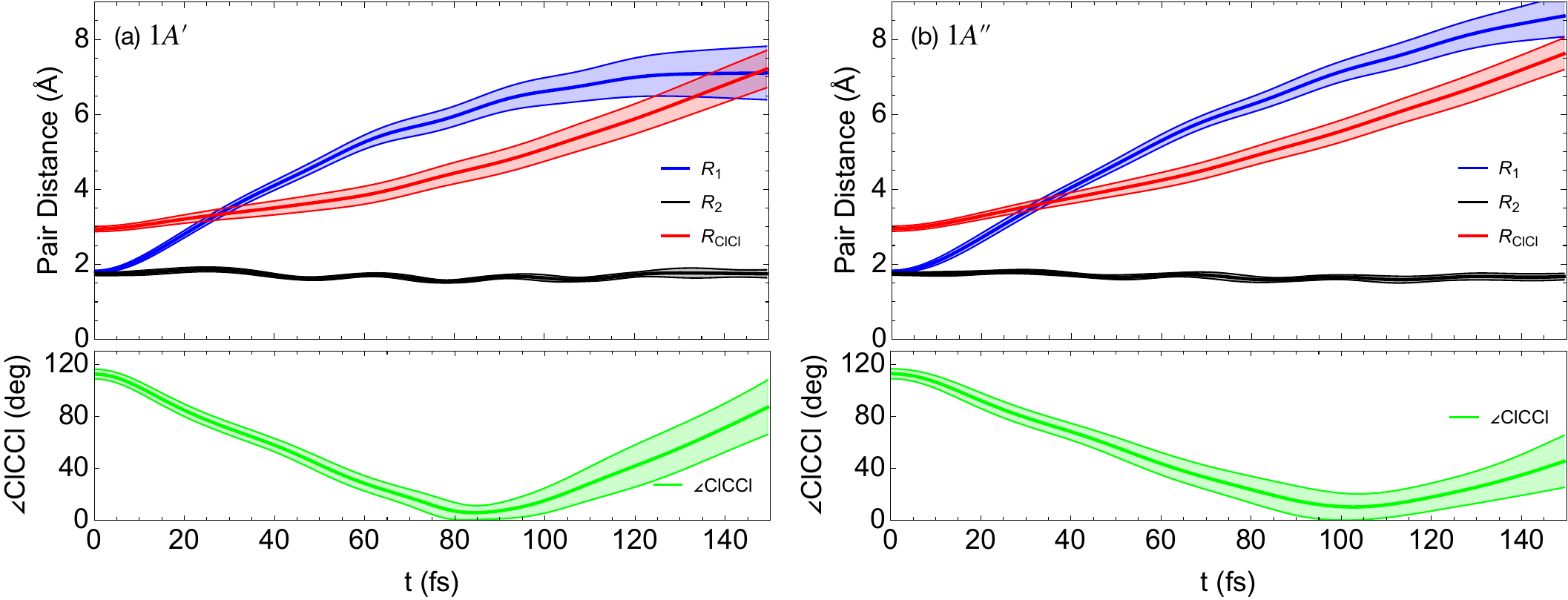}
  \caption{The three pair distances C-Cl(1) (blue, dissociation coordinate), C-Cl(2) (black), Cl-Cl (red), and the ClCCl angle (green), as a function of propagation time for 1000 trajectories on the $1A'$ (a) and $1A''$ (b) PESs, respectively. The thick solid curves represent the average values and the shadow areas indicate the average values $\pm$ the standard deviation for the 1000 trajectories. }
  \label{fig4}
\end{figure}

In the UED imaging method, inter-nuclear distances are typically derived from the electron diffraction pattern and used to reconstruct the molecular structure. Figure \ref{fig4} shows the pair distances as a function of propagation time for 1000 adiabatic trajectories on the 1$A'$ and 1$A''$ states, respectively.
In the case of two-body dissociation, one C–Cl bond length remains almost unchanged, exhibiting only small vibrations. Meanwhile, the Cl–Cl distance increases significantly and even accelerates over time. Molecules on the 1$A''$ state dissociate slightly faster than those on the 1$A'$ state.
During photodissociation, the $\text{CH}_2\text{Cl}$ rotation is also excited, with an average period of approximately 300 fs. As a result, the C–Cl dissociation distance does not increase monotonically. Instead, the C–Cl pair distance saturates around 120 fs on the 1$A'$ state and then decreases slightly as the Cl atom dissociates.

\subsection{Simulating observables in Coulomb explosion experiments}
In this work, we focus on studying the three-fragment Coulomb explosion in the lowest charged channel, specifically $\text{CH}_2^+ + \text{Cl}^+ + \text{Cl}^+$ fragmentation. When using tabletop lasers, this triply charged state is typically achieved through strong-field-induced multifold ionization.
In a typical cold target recoil-ion momentum (COLTRIM) experiment, the asymptotic momenta of these three ionic fragments are measured in coincidence to reconstruct the time-resolved molecular structure\cite{endo2020,ziaee2023,dorner2000}. Since the $\text{CH}_2^+$ fragment is detected as a whole, the internal structure of $\text{CH}_2$ is not resolved.
From a theoretical perspective, developing rigorous theories for strong-field-induced Coulomb explosion is challenging. First, modeling strong-field multifold ionization in polyatomic molecules is difficult. Second, interactions between ionic fragments can be intricate, particularly when the fragments are in close proximity or when the exact electronic states of the multicharged ion are unknown.

In our MD simulation, we neglect the pulse duration and its effect on nuclear dynamics, assuming that photoionization occurs instantaneously. Consequently, the trajectory propagation time on the neutral PES is treated as the pump-probe delay. We also assume the interaction between ionic fragments to be purely coulombic, which can be written as
\begin{equation}
    V_{CE}(R_1,R_2,\alpha)=\frac{1}{R_1}+\frac{1}{R_2}+\frac{1}{\sqrt{R_1^2+R_2^2-2R_1R_2\cos(2\alpha)}}.
    \label{eqcoulomb}
\end{equation}
The Coulomb explosion simulation is performed by solving the same EOM as for the neutral molecule (i.e., Eq. \eqref{eqeom}), with the PES replaced by the pure Coulomb potential Eq. \eqref{eqcoulomb}. We use the $\text{CH}_2\text{Cl}_2$ trajectory state at a specific pump-probe delay as the initial condition and propagate the trajectory for $\text{CH}_2\text{Cl}_2^{3+}$ ions until $V_{CE}<0.05$ eV. 

The kinetic energy release (KER) is a key observable in CEI experiments. It is the sum of the translational energies of $\text{Cl}^+$ , $\text{Cl}^+$, and $\text{CH}_2^+$ fragments detected in coincidence. With knowledge of the asymptotic momenta $\textbf{p}_i(i=\text{CH}_2^+,\text{Cl}^+,\text{Cl}^+)$, the KER can be written as 
\begin{equation}
    \mathrm{KER}=\sum_{i=\text{CH}_2^+,\text{Cl}^+,\text{Cl}^+}\frac{\textbf{p}_i^2}{2m_i}.
    \label{eqkermd}
\end{equation}
Without simulating the Coulomb explosion dynamics, the KER can still be estimated from energy conservation by assuming the $\text{CH}_2\text{Cl}_2^{3+}$ potential energy completely converted to the kinetic energies of the ionic fragments. In this case, the KER can be approximated as
\begin{equation}
    \mathrm{KER}\approx\sum_{i=\text{CH}_2,\text{Cl},\text{Cl}}\frac{\textbf{p}_i^2}{2m_i}+V_{CE}(R_1,R_2,\alpha),
    \label{eqkerce}
\end{equation}
where $\textbf{p}_i(i=\text{CH}_2,\text{Cl},\text{Cl})$ are the momenta of the neutral components at a certain pump-probe delay. Since the actual $\text{CH}_2\text{Cl}_2^{3+}$ potential is not purely Coulombic, this estimation can be improved using \textit{ab initio} potential energies of $\text{CH}_2\text{Cl}_2^{3+}$ (denoted by $V_{ion}$), that is, 
\begin{equation}
    \mathrm{KER}\approx\sum_{i=\text{CH}_2,\text{Cl},\text{Cl}}\frac{\textbf{p}_i^2}{2m_i}+V_{ion}(R_1,R_2,\alpha),
    \label{eqkermulti}
\end{equation}
where $V_{ion}$ is calculated using the multi-configuration self-consistent field (MCSCF) method\cite{mcscf1,mcscf2} with an active space of 11 electrons in 10 orbitals and the 6-311G** basis set\cite{krishnan1980a}. 

\begin{figure}
  \includegraphics[width=0.8\textwidth]{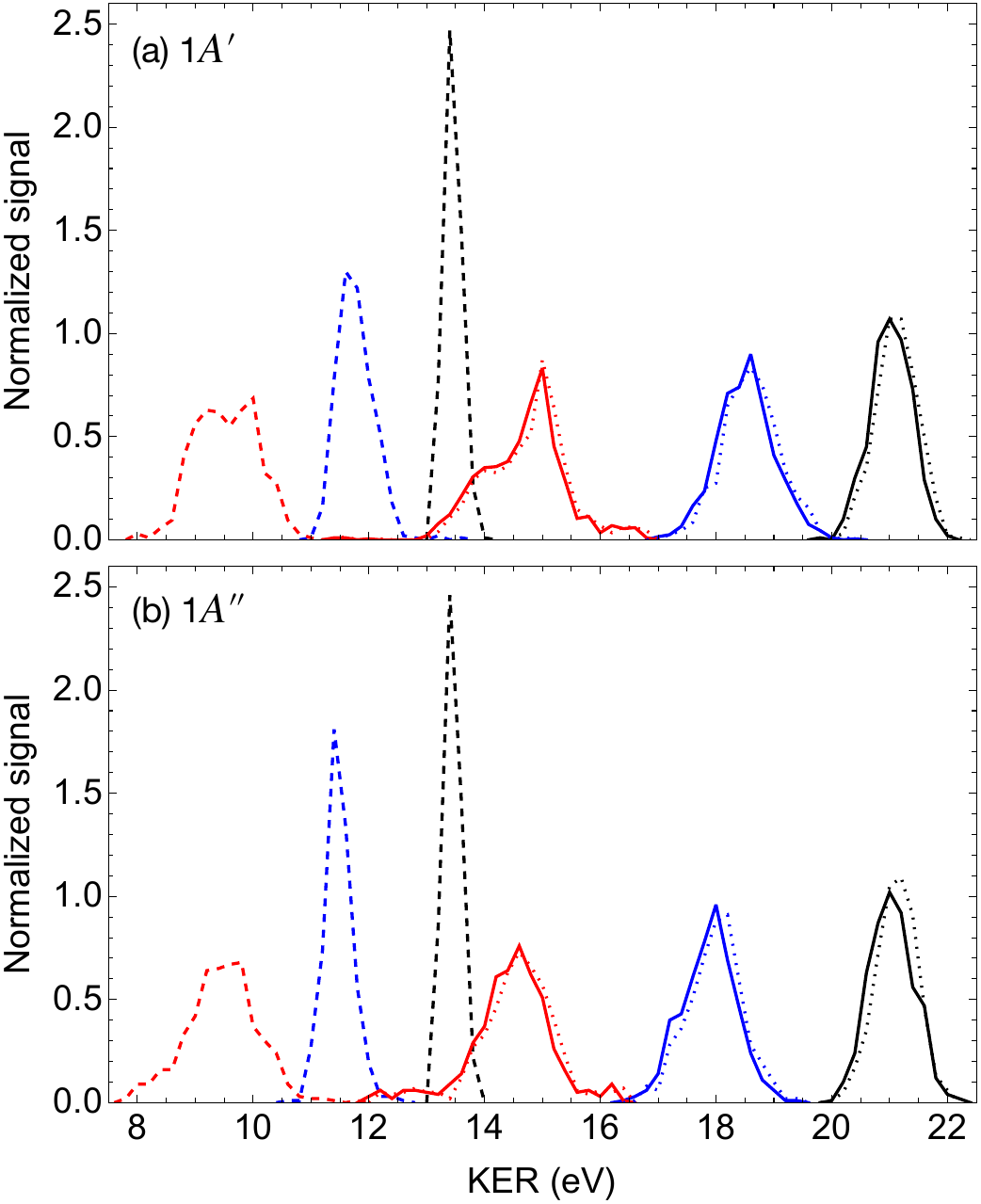}
  \caption{The kinetic energy release (KER) distribution for signals on 1$A'$ (a) and 1$A''$ (b) dissociative states, respectively, for pump-probe delays at 0 fs (black), 50 fs (blue), and 150 fs (red), where the Coulomb explosion occurs in the $\text{CH}_2^+ +\text{Cl}^+ + \text{Cl}^+$ fragmentation channel. The solid curves indicate KER obtained by simulating the Coulomb explosion dynamics using the Coulomb potential Eq. \eqref{eqcoulomb}. The dashed curves indicate KER estimated using \textit{ab initio} potential energies of $\text{CH}_2\text{Cl}_2^{3+}$, calculated with the MCSCF(11,10)/6-311G** method, as denoted by Eq. \eqref{eqkermulti}.
  The dotted curves indicate KER estimated using the pure Coulomb potential, as denoted by Eq. \eqref{eqkerce}. Both the dashed and dotted curves assume that ionic potential energies are completely converted into KER.
  }
  \label{fig5}
\end{figure}

Figure \ref{fig5} shows the KER distribution calculated using different approaches for pump-probe delays of 0 fs, 50 fs, and 150 fs, respectively. The time-resolved KER signals decrease significantly with increasing pump-probe delay due to the $\text{CH}_2\text{Cl} + \text{Cl}$ dissociation process.
The KER obtained from the dynamics simulation closely matches the estimation from Eq. \eqref{eqkerce}, indicating that the ionic potential energy is largely converted into translational energies, with only a minimal fraction redistributed to $\text{CH}_2^+$ rotation. 
The KER estimated from $\text{CH}_2\text{Cl}_2^{3+}$ \textit{ab initio} calculations is significantly lower than the values obtained using a pure Coulomb potential for all pump-probe delays, which attributes to the covalent attraction between the ionic fragments. The difference is approximately 8 eV at 0 fs, and decreases to about 5 eV at 150 fs, as the interaction becomes more Coulombic with increasing C-Cl and Cl-Cl distances.
In fact, the KER calculated using the \textit{ab initio} ionic potential is expected to align more closely with experimental results, as demonstrated in studies of other halomethane species\cite{ziaee2023,ding2024,corrales2012}.

The time-resolved coincident fragmentation signals in the COLTRIM experiment offer us abundant information about the transient structural changes of the molecule. We analyze the $\text{CH}_2^+ +\text{Cl}^+ + \text{Cl}^+$ coincidence signals as a function of KER and the angle between the momentum vectors of the two $\text{Cl}^+$ fragments, that is, 
\begin{equation}
    \cos\Theta=\frac{\textbf{p}_{\text{Cl}^+(1)}\cdot \textbf{p}_{\text{Cl}^+(2)}}{|\textbf{p}_{\text{Cl}^+(1)}||\textbf{p}_{\text{Cl}^+(2)}|}
\end{equation}
Similar analysis is also performed on other molecules\cite{liekhus2015,balramthesis}.

\begin{figure}
  \includegraphics[width=\textwidth]{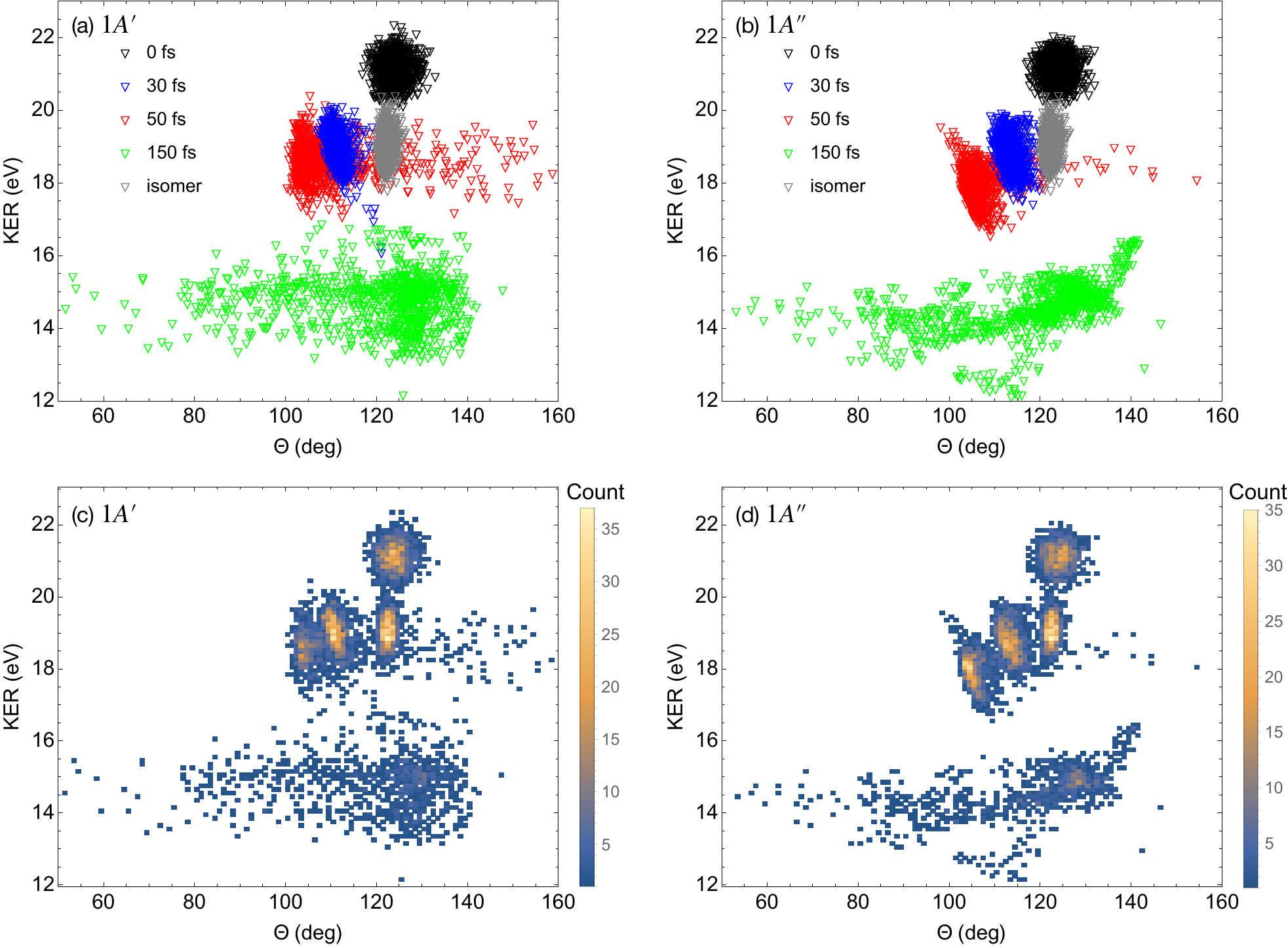}
  \caption{The simulated $\text{CH}_2^+ +\text{Cl}^+ + \text{Cl}^+$ coincidence signals as a function of the kinetic energy release (KER) and the angle $\Theta$ between the final momentum vectors of the two $\text{Cl}^+$ fragments for pump-probe delays at 0 fs, 30 fs, 50 fs, and 150 fs. The simulated isomer signals, sampled from the Wigner distribution, are also shown for comparison. (a) and (b) are simulated signals directly shown as a function of KER and $\Theta$. The photodissociation occurs on 1$A'$ (a) and 1$A''$ (b) states. (c) and (d) are density histograms that count all signals in (a) and (b), using a bin size of $\Delta E=0.1$ eV and $\Delta\Theta=1^\circ$.
  }
  \label{fig6}
\end{figure}

The results are shown in Fig. \ref{fig6}. As the pump-probe delay increases, the KER signals decrease significantly due to C-Cl bond breaking. Meanwhile, the angular distribution of $\Theta$ also spreads due to the excitation of $\text{CH}_2\text{Cl}$ rotational motion: at 50 fs, $\Theta$ extends to 160 degree and at 150 fs, $\Theta$ extends to about 50 degree. Such a spreading is more significant in the 1$A'$ state than in the 1$A''$ state. 

We also investigate whether the unlikelihood of isomerization will be reflected in the CEI coincidence signals. To this end, we also simulate the Coulomb explosion for nuclear geometries near the isomer. 
The three normal frequencies near the isomer geometry are $\{\omega_1,\omega_2,\omega_3\}=\{181,305,1050\}$ cm$^{-1}$. We sample 1000 initial conditions using the Wigner distribution Eq.~\eqref{eqwigner} at T=300 K, and simulate the Coulomb explosion. The results are also shown in Fig. \ref{fig6} for comparison. The time-resolved signals for photodissociation dynamics are separated from that of the isomer for both 1$A'$ and 1$A''$ states , which confirms the unlikelihood of intra-molecular isomerization in $\text{CH}_2\text{Cl}_2$. 

\section{Conclusion}
In summary, we have performed molecular dynamics simulations of $\text{CH}_2\text{Cl}_2$ photodissociation on its VUV-excited states and the subsequent Coulomb explosion on the triply charged state. We have constructed three-dimensional potential energy surfaces for the low-lying excited states of $\text{CH}_2\text{Cl}_2$, capable of characterizing C–Cl bond breaking and rearrangement processes, based on high-level \textit{ab initio} calculations and a tri-cubic spline interpolation algorithm.
Trajectory calculations for the first two excited states indicate that intramolecular photoisomerization of $\text{CH}_2\text{Cl}_2$ is unlikely to occur. Most trajectories exhibit direct two-body dissociation, producing $\text{CH}_2\text{Cl}$ and Cl radical products while inducing rotational motion in $\text{CH}_2\text{Cl}$. 
As a key observable in Coulomb explosion experiments, the time-resolved KER is calculated from molecular dynamics simulations and is also estimated from energy conservation principles. The KER estimated using \textit{ab initio} ionic potential energies, which is expected to be more realistic, is approximately 5–8 eV lower than that estimated using pure Coulomb potential.
The molecular structural changes during photodissociation are mapped onto time-resolved $\text{CH}_2^+ + \text{Cl}^+ + \text{Cl}^+$ coincidence signals, which are analyzed in terms of KER and the angle between the two $\text{Cl}^+$ recoil momenta. These coincidence signals further confirm the unlikelihood of photoisomerization in $\text{CH}_2\text{Cl}_2$ and reveal the rotational excitation of the $\text{CH}_2\text{Cl}$ radical during photodissociation.

Our theoretical work provides valuable insights into the photodissociation dynamics of $\text{CH}_2\text{Cl}_2$ at the molecular level, which has not been investigated previously. The simulated observables offer useful guidance for ultrafast experiments on this molecule. The corresponding pump-probe experiment in our lab is also on the schedule\cite{privatecomm}. 
%We hope our theoretical work will stimulate more relevant studies on this molecule. 

%%%%%%%%%%%%%%%%%%%%%%%%%%%%%%%%%%%%%%%%%%%%%%%%%%%%%%%%%%%%%%%%%%%%%%%%
%% The "Acknowledgement" section can be given in all manuscript       %%
%% classes.  This should be given within the "acknowledgement"        %%
%% environment, which will make the correct section or running title. %%
%%%%%%%%%%%%%%%%%%%%%%%%%%%%%%%%%%%%%%%%%%%%%%%%%%%%%%%%%%%%%%%%%%%%%%%%
\begin{acknowledgement}

The author thanks Prof. Daniel Rolles for fruitful discussions from the experimental perspective and for careful reading of the manuscript. The author thanks Panwang Zhou for discussions on the \textit{ab initio} calculations. The author thanks Kansas State University for providing computing resources to generate certain results in this work.
%This work used resources of Beocat high-performance computing cluster at Kansas State University.

\end{acknowledgement}

%\begin{suppinfo}

%This will usually read something like: ``Experimental procedures and
%characterization data for all new compounds. The class will
%automatically add a sentence pointing to the information on-line:

%The potential energy surfaces are available upon request to the corresponding author.

%\end{suppinfo}

\bibliography{achemso-demo}

\providecommand{\latin}[1]{#1}
\makeatletter
\providecommand{\doi}
  {\begingroup\let\do\@makeother\dospecials
  \catcode`\{=1 \catcode`\}=2 \doi@aux}
\providecommand{\doi@aux}[1]{\endgroup\texttt{#1}}
\makeatother
\providecommand*\mcitethebibliography{\thebibliography}
\csname @ifundefined\endcsname{endmcitethebibliography}
  {\let\endmcitethebibliography\endthebibliography}{}
\begin{mcitethebibliography}{43}
\providecommand*\natexlab[1]{#1}
\providecommand*\mciteSetBstSublistMode[1]{}
\providecommand*\mciteSetBstMaxWidthForm[2]{}
\providecommand*\mciteBstWouldAddEndPuncttrue
  {\def\EndOfBibitem{\unskip.}}
\providecommand*\mciteBstWouldAddEndPunctfalse
  {\let\EndOfBibitem\relax}
\providecommand*\mciteSetBstMidEndSepPunct[3]{}
\providecommand*\mciteSetBstSublistLabelBeginEnd[3]{}
\providecommand*\EndOfBibitem{}
\mciteSetBstSublistMode{f}
\mciteSetBstMaxWidthForm{subitem}{(\alph{mcitesubitemcount})}
\mciteSetBstSublistLabelBeginEnd
  {\mcitemaxwidthsubitemform\space}
  {\relax}
  {\relax}

\bibitem[Simmonds \latin{et~al.}(2006)Simmonds, Manning, Cunnold, McCulloch,
  O'Doherty, Derwent, Krummel, Fraser, Dunse, Porter, Wang, Greally, Miller,
  Salameh, Weiss, and Prinn]{simmonds2006}
Simmonds,~P.~G. \latin{et~al.}  Global trends, seasonal cycles, and European
  emissions of dichloromethane, trichloroethene, and tetrachloroethene from the
  AGAGE observations at Mace Head, Ireland, and Cape Grim, Tasmania.
  \emph{Journal of Geophysical Research: Atmospheres} \textbf{2006},
  \emph{111}, D18304\relax
\mciteBstWouldAddEndPuncttrue
\mciteSetBstMidEndSepPunct{\mcitedefaultmidpunct}
{\mcitedefaultendpunct}{\mcitedefaultseppunct}\relax
\EndOfBibitem
\bibitem[Hossaini \latin{et~al.}(2017)Hossaini, Chipperfield, Montzka, Leeson,
  Dhomse, and Pyle]{hossaini2017}
Hossaini,~R.; Chipperfield,~M.~P.; Montzka,~S.~A.; Leeson,~A.~A.;
  Dhomse,~S.~S.; Pyle,~J.~A. The increasing threat to stratospheric ozone from
  dichloromethane. \emph{Nature Communications} \textbf{2017}, \emph{8},
  15962\relax
\mciteBstWouldAddEndPuncttrue
\mciteSetBstMidEndSepPunct{\mcitedefaultmidpunct}
{\mcitedefaultendpunct}{\mcitedefaultseppunct}\relax
\EndOfBibitem
\bibitem[Mandal \latin{et~al.}(2014)Mandal, Singh, Shastri, and
  Jagatap]{mandal2014}
Mandal,~A.; Singh,~P.~J.; Shastri,~A.; Jagatap,~B. Vacuum ultraviolet
  photoabsorption spectroscopy of CH2Cl2 and CD2Cl2 in the energy region
  50,000-95,000cm-1. \emph{Journal of Quantitative Spectroscopy and Radiative
  Transfer} \textbf{2014}, \emph{149}, 291--302\relax
\mciteBstWouldAddEndPuncttrue
\mciteSetBstMidEndSepPunct{\mcitedefaultmidpunct}
{\mcitedefaultendpunct}{\mcitedefaultseppunct}\relax
\EndOfBibitem
\bibitem[Lange \latin{et~al.}(2020)Lange, Jones, Hoffmann, Lozano, Kumar,
  Homem, Śmiałek, Duflot, Brunger, and Limão-Vieira]{lange2020}
Lange,~E.; Jones,~N.; Hoffmann,~S.; Lozano,~A.; Kumar,~S.; Homem,~M.;
  Śmiałek,~M.; Duflot,~D.; Brunger,~M.; Limão-Vieira,~P. The electronic
  excited states of dichloromethane in the 5.8-10.8 eV energy range
  investigated by experimental and theoretical methods. \emph{Journal of
  Quantitative Spectroscopy and Radiative Transfer} \textbf{2020}, \emph{253},
  107172\relax
\mciteBstWouldAddEndPuncttrue
\mciteSetBstMidEndSepPunct{\mcitedefaultmidpunct}
{\mcitedefaultendpunct}{\mcitedefaultseppunct}\relax
\EndOfBibitem
\bibitem[Vager \latin{et~al.}(1989)Vager, Naaman, and Kanter]{cei1}
Vager,~Z.; Naaman,~R.; Kanter,~E.~P. Coulomb Explosion Imaging of Small
  Molecules. \emph{Science} \textbf{1989}, \emph{244}, 426--431\relax
\mciteBstWouldAddEndPuncttrue
\mciteSetBstMidEndSepPunct{\mcitedefaultmidpunct}
{\mcitedefaultendpunct}{\mcitedefaultseppunct}\relax
\EndOfBibitem
\bibitem[Schouder \latin{et~al.}(2022)Schouder, Chatterley, Pickering, and
  Stapelfeldt]{cei2}
Schouder,~C.~A.; Chatterley,~A.~S.; Pickering,~J.~D.; Stapelfeldt,~H.
  Laser-Induced Coulomb Explosion Imaging of Aligned Molecules and Molecular
  Dimers. \emph{Annual Review of Physical Chemistry} \textbf{2022}, \emph{73},
  323--347\relax
\mciteBstWouldAddEndPuncttrue
\mciteSetBstMidEndSepPunct{\mcitedefaultmidpunct}
{\mcitedefaultendpunct}{\mcitedefaultseppunct}\relax
\EndOfBibitem
\bibitem[Filippetto \latin{et~al.}(2022)Filippetto, Musumeci, Li, Siwick, Otto,
  Centurion, and Nunes]{ued}
Filippetto,~D.; Musumeci,~P.; Li,~R.~K.; Siwick,~B.~J.; Otto,~M.~R.;
  Centurion,~M.; Nunes,~J. P.~F. Ultrafast electron diffraction: Visualizing
  dynamic states of matter. \emph{Rev. Mod. Phys.} \textbf{2022}, \emph{94},
  045004\relax
\mciteBstWouldAddEndPuncttrue
\mciteSetBstMidEndSepPunct{\mcitedefaultmidpunct}
{\mcitedefaultendpunct}{\mcitedefaultseppunct}\relax
\EndOfBibitem
\bibitem[Ziaee \latin{et~al.}(2023)Ziaee, Borne, Forbes, P., Malakar, Kaderiya,
  Severt, Ben-Itzhak, Rudenko, and Rolles]{ziaee2023}
Ziaee,~F.; Borne,~K.; Forbes,~R.; P.,~K.~R.; Malakar,~Y.; Kaderiya,~B.;
  Severt,~T.; Ben-Itzhak,~I.; Rudenko,~A.; Rolles,~D. Single- and
  multi-photon-induced ultraviolet excitation and photodissociation of CH3I
  probed by coincident ion momentum imaging. \emph{Phys. Chem. Chem. Phys.}
  \textbf{2023}, \emph{25}, 9999--10010\relax
\mciteBstWouldAddEndPuncttrue
\mciteSetBstMidEndSepPunct{\mcitedefaultmidpunct}
{\mcitedefaultendpunct}{\mcitedefaultseppunct}\relax
\EndOfBibitem
\bibitem[Ding \latin{et~al.}(2024)Ding, Greenman, and Rolles]{ding2024}
Ding,~Y.; Greenman,~L.; Rolles,~D. Surface hopping molecular dynamics
  simulation of ultrafast methyl iodide photodissociation mapped by Coulomb
  explosion imaging. \emph{Phys. Chem. Chem. Phys.} \textbf{2024}, \emph{26},
  22423--22432\relax
\mciteBstWouldAddEndPuncttrue
\mciteSetBstMidEndSepPunct{\mcitedefaultmidpunct}
{\mcitedefaultendpunct}{\mcitedefaultseppunct}\relax
\EndOfBibitem
\bibitem[Bhattacharyya \latin{et~al.}(2024)Bhattacharyya, Wang, Borne, Chen,
  Venkatachalam, Lam, Ziaee, Pathak, Khmelnitskiy, Carnes, Fehrenbach,
  Ben-Itzhak, Rudenko, and Rolles]{surgendu2024}
Bhattacharyya,~S.; Wang,~E.; Borne,~K.; Chen,~K.; Venkatachalam,~A.~S.; Lam,~H.
  V.~S.; Ziaee,~F.; Pathak,~S.; Khmelnitskiy,~A.; Carnes,~K.~D.;
  Fehrenbach,~C.~W.; Ben-Itzhak,~I.; Rudenko,~A.; Rolles,~D. Delayed
  Dissociation and Transient Isomerization during the Ultrafast
  Photodissociation of the Tribromomethane Cation. \emph{The Journal of
  Physical Chemistry Letters} \textbf{2024}, \emph{15}, 12188--12196\relax
\mciteBstWouldAddEndPuncttrue
\mciteSetBstMidEndSepPunct{\mcitedefaultmidpunct}
{\mcitedefaultendpunct}{\mcitedefaultseppunct}\relax
\EndOfBibitem
\bibitem[Endo \latin{et~al.}(2020)Endo, Neville, Wanie, Beaulieu, Qu,
  Deschamps, Lassonde, Schmidt, Fujise, Fushitani, Hishikawa, Houston, Bowman,
  Schuurman, Légaré, and Ibrahim]{endo2020}
Endo,~T. \latin{et~al.}  Capturing roaming molecular fragments in real time.
  \emph{Science} \textbf{2020}, \emph{370}, 1072--1077\relax
\mciteBstWouldAddEndPuncttrue
\mciteSetBstMidEndSepPunct{\mcitedefaultmidpunct}
{\mcitedefaultendpunct}{\mcitedefaultseppunct}\relax
\EndOfBibitem
\bibitem[Li \latin{et~al.}(2021)Li, Zhang, Vendrell, Guo, Zhu, Gao, Cao, Guo,
  Su, Cao, Luo, Yan, Zhou, Liu, Li, and Lu]{li2021}
Li,~M. \latin{et~al.}  Ultrafast imaging of spontaneous symmetry breaking in a
  photoionized molecular system. \emph{Nat. Commun.} \textbf{2021}, \emph{12},
  4233\relax
\mciteBstWouldAddEndPuncttrue
\mciteSetBstMidEndSepPunct{\mcitedefaultmidpunct}
{\mcitedefaultendpunct}{\mcitedefaultseppunct}\relax
\EndOfBibitem
\bibitem[Wang \latin{et~al.}(2024)Wang, Dong, Zhang, Deng, Jian, Li, and
  Liu]{wang2024}
Wang,~J.; Dong,~B.; Zhang,~M.; Deng,~Y.; Jian,~X.; Li,~Z.; Liu,~Y. Ultrafast
  Imaging of Jahn--Teller Distortion and the Correlated Proton Migration in
  Photoionized Cyclopropane. \emph{J. Am. Chem. Soc.} \textbf{2024},
  \emph{146}, 10443--10450\relax
\mciteBstWouldAddEndPuncttrue
\mciteSetBstMidEndSepPunct{\mcitedefaultmidpunct}
{\mcitedefaultendpunct}{\mcitedefaultseppunct}\relax
\EndOfBibitem
\bibitem[Crespo-Hernández \latin{et~al.}(2004)Crespo-Hernández, Cohen, Hare,
  and Kohler]{carlos2004}
Crespo-Hernández,~C.~E.; Cohen,~B.; Hare,~P.~M.; Kohler,~B. Ultrafast
  Excited-State Dynamics in Nucleic Acids. \emph{Chemical Reviews}
  \textbf{2004}, \emph{104}, 1977--2020\relax
\mciteBstWouldAddEndPuncttrue
\mciteSetBstMidEndSepPunct{\mcitedefaultmidpunct}
{\mcitedefaultendpunct}{\mcitedefaultseppunct}\relax
\EndOfBibitem
\bibitem[Polli \latin{et~al.}(2010)Polli, Alto{\`e}, Weingart, Spillane,
  Manzoni, Brida, Tomasello, Orlandi, Kukura, Mathies, Garavelli, and
  Cerullo]{polli2010}
Polli,~D.; Alto{\`e},~P.; Weingart,~O.; Spillane,~K.~M.; Manzoni,~C.;
  Brida,~D.; Tomasello,~G.; Orlandi,~G.; Kukura,~P.; Mathies,~R.~A.;
  Garavelli,~M.; Cerullo,~G. Conical intersection dynamics of the primary
  photoisomerization event in vision. \emph{Nature} \textbf{2010}, \emph{467},
  440--443\relax
\mciteBstWouldAddEndPuncttrue
\mciteSetBstMidEndSepPunct{\mcitedefaultmidpunct}
{\mcitedefaultendpunct}{\mcitedefaultseppunct}\relax
\EndOfBibitem
\bibitem[Ding \latin{et~al.}(2016)Ding, Pérez-Ríos, and Greene]{ding2016}
Ding,~Y.; Pérez-Ríos,~J.; Greene,~C.~H. Effective Atom–Molecule Conversions
  Using Radio Frequency Fields. \emph{ChemPhysChem} \textbf{2016}, \emph{17},
  3756--3763\relax
\mciteBstWouldAddEndPuncttrue
\mciteSetBstMidEndSepPunct{\mcitedefaultmidpunct}
{\mcitedefaultendpunct}{\mcitedefaultseppunct}\relax
\EndOfBibitem
\bibitem[Ding \latin{et~al.}(2017)Ding, D'Incao, and Greene]{ding2017}
Ding,~Y.; D'Incao,~J.~P.; Greene,~C.~H. Effective control of cold collisions
  with radio-frequency fields. \emph{Phys. Rev. A} \textbf{2017}, \emph{95},
  022709\relax
\mciteBstWouldAddEndPuncttrue
\mciteSetBstMidEndSepPunct{\mcitedefaultmidpunct}
{\mcitedefaultendpunct}{\mcitedefaultseppunct}\relax
\EndOfBibitem
\bibitem[Xiao \latin{et~al.}(2007)Xiao, Liu, Yu, and Fang]{xiao2007}
Xiao,~H.-Y.; Liu,~Y.-J.; Yu,~J.-G.; Fang,~W.-H. Spin–orbit ab initio
  investigation of the photodissociation of CH2Cl2. \emph{Chemical Physics
  Letters} \textbf{2007}, \emph{436}, 75--79\relax
\mciteBstWouldAddEndPuncttrue
\mciteSetBstMidEndSepPunct{\mcitedefaultmidpunct}
{\mcitedefaultendpunct}{\mcitedefaultseppunct}\relax
\EndOfBibitem
\bibitem[Almásy and Bende(2019)Almásy, and Bende]{almasy2019}
Almásy,~L.; Bende,~A. Intermolecular Interaction in Methylene Halide (CH2F2,
  CH2Cl2, CH2Br2 and CH2I2) Dimers. \emph{Molecules} \textbf{2019}, \emph{24},
  1810\relax
\mciteBstWouldAddEndPuncttrue
\mciteSetBstMidEndSepPunct{\mcitedefaultmidpunct}
{\mcitedefaultendpunct}{\mcitedefaultseppunct}\relax
\EndOfBibitem
\bibitem[Lewars(1998)]{lewars1998}
Lewars,~E. The isomers of dichloromethane and its radical cation: an ab initio
  exploration of the neutral and charged CH2Cl2 potential energy surfaces.
  \emph{Journal of Molecular Structure: THEOCHEM} \textbf{1998}, \emph{425},
  207--226\relax
\mciteBstWouldAddEndPuncttrue
\mciteSetBstMidEndSepPunct{\mcitedefaultmidpunct}
{\mcitedefaultendpunct}{\mcitedefaultseppunct}\relax
\EndOfBibitem
\bibitem[Tonokura \latin{et~al.}(1992)Tonokura, Mo, Matsumi, and
  Kawasaki]{tonokura1992}
Tonokura,~K.; Mo,~Y.; Matsumi,~Y.; Kawasaki,~M. Doppler spectroscopy of
  hydrogen and chlorine atoms from photodissociation of silane, germane,
  chlorosilanes, and chloromethanes in the vacuum ultraviolet region. \emph{The
  Journal of Physical Chemistry} \textbf{1992}, \emph{96}, 6688--6693\relax
\mciteBstWouldAddEndPuncttrue
\mciteSetBstMidEndSepPunct{\mcitedefaultmidpunct}
{\mcitedefaultendpunct}{\mcitedefaultseppunct}\relax
\EndOfBibitem
\bibitem[Brownsword \latin{et~al.}(1997)Brownsword, Hillenkamp, Laurent, Vatsa,
  Volpp, and Wolfrum]{brownsword1997}
Brownsword,~R.~A.; Hillenkamp,~M.; Laurent,~T.; Vatsa,~R.~K.; Volpp,~H.-R.;
  Wolfrum,~J. Photodissociation dynamics of the chloromethanes at the Lyman-a
  wavelength (121.6 nm). \emph{The Journal of Chemical Physics} \textbf{1997},
  \emph{106}, 1359--1366\relax
\mciteBstWouldAddEndPuncttrue
\mciteSetBstMidEndSepPunct{\mcitedefaultmidpunct}
{\mcitedefaultendpunct}{\mcitedefaultseppunct}\relax
\EndOfBibitem
\bibitem[Hayakawa \latin{et~al.}(2008)Hayakawa, Sasaki, and
  Matsubara]{hayakawa2008}
Hayakawa,~S.; Sasaki,~T.; Matsubara,~H. Dissociation mechanism of
  electronically excited CH2X2 (X=Cl, Br) formed by near-resonant
  neutralization using charge-inversion mass spectrometry. \emph{Chemical
  Physics Letters} \textbf{2008}, \emph{463}, 60--64\relax
\mciteBstWouldAddEndPuncttrue
\mciteSetBstMidEndSepPunct{\mcitedefaultmidpunct}
{\mcitedefaultendpunct}{\mcitedefaultseppunct}\relax
\EndOfBibitem
\bibitem[Gagnon \latin{et~al.}(2008)Gagnon, Lee, Rayner, Corkum, and
  Bhardwaj]{gagnon2008}
Gagnon,~J.; Lee,~K.~F.; Rayner,~D.~M.; Corkum,~P.~B.; Bhardwaj,~V.~R.
  Coincidence imaging of polyatomic molecules via laser-induced Coulomb
  explosion. \emph{Journal of Physics B: Atomic, Molecular and Optical Physics}
  \textbf{2008}, \emph{41}, 215104\relax
\mciteBstWouldAddEndPuncttrue
\mciteSetBstMidEndSepPunct{\mcitedefaultmidpunct}
{\mcitedefaultendpunct}{\mcitedefaultseppunct}\relax
\EndOfBibitem
\bibitem[pri()]{privatecomm}
Private communication with Prof. Daniel Rolles. \relax
\mciteBstWouldAddEndPunctfalse
\mciteSetBstMidEndSepPunct{\mcitedefaultmidpunct}
{}{\mcitedefaultseppunct}\relax
\EndOfBibitem
\bibitem[Hampel \latin{et~al.}(1992)Hampel, Peterson, and Werner]{hampel1992}
Hampel,~C.; Peterson,~K.~A.; Werner,~H.-J. A comparison of the efficiency and
  accuracy of the quadratic configuration interaction (QCISD), coupled cluster
  (CCSD), and Brueckner coupled cluster (BCCD) methods. \emph{Chemical Physics
  Letters} \textbf{1992}, \emph{190}, 1--12\relax
\mciteBstWouldAddEndPuncttrue
\mciteSetBstMidEndSepPunct{\mcitedefaultmidpunct}
{\mcitedefaultendpunct}{\mcitedefaultseppunct}\relax
\EndOfBibitem
\bibitem[Korona and Werner(2003)Korona, and Werner]{eomccsd}
Korona,~T.; Werner,~H.-J. {Local treatment of electron excitations in the
  EOM-CCSD method}. \emph{The Journal of Chemical Physics} \textbf{2003},
  \emph{118}, 3006--3019\relax
\mciteBstWouldAddEndPuncttrue
\mciteSetBstMidEndSepPunct{\mcitedefaultmidpunct}
{\mcitedefaultendpunct}{\mcitedefaultseppunct}\relax
\EndOfBibitem
\bibitem[Werner \latin{et~al.}(2020)Werner, Knowles, Manby, Black, Doll,
  Heßelmann, Kats, Köhn, Korona, Kreplin, Ma, Miller, Mitrushchenkov,
  Peterson, Polyak, Rauhut, and Sibaev]{molpro}
Werner,~H.-J. \latin{et~al.}  {The Molpro quantum chemistry package}. \emph{J.
  Chem. Phys.} \textbf{2020}, \emph{152}, 144107\relax
\mciteBstWouldAddEndPuncttrue
\mciteSetBstMidEndSepPunct{\mcitedefaultmidpunct}
{\mcitedefaultendpunct}{\mcitedefaultseppunct}\relax
\EndOfBibitem
\bibitem[Werner \latin{et~al.}(2012)Werner, Knowles, Knizia, Manby, and
  Schütz]{molpro2}
Werner,~H.-J.; Knowles,~P.~J.; Knizia,~G.; Manby,~F.~R.; Schütz,~M. Molpro: a
  general-purpose quantum chemistry program package. \emph{WIREs Computational
  Molecular Science} \textbf{2012}, \emph{2}, 242--253\relax
\mciteBstWouldAddEndPuncttrue
\mciteSetBstMidEndSepPunct{\mcitedefaultmidpunct}
{\mcitedefaultendpunct}{\mcitedefaultseppunct}\relax
\EndOfBibitem
\bibitem[Dunning(1989)]{dunningbasis}
Dunning,~J.,~Thom~H. {Gaussian basis sets for use in correlated molecular
  calculations. I. The atoms boron through neon and hydrogen}. \emph{J. Chem.
  Phys.} \textbf{1989}, \emph{90}, 1007--1023\relax
\mciteBstWouldAddEndPuncttrue
\mciteSetBstMidEndSepPunct{\mcitedefaultmidpunct}
{\mcitedefaultendpunct}{\mcitedefaultseppunct}\relax
\EndOfBibitem
\bibitem[Uehara and Horiai(1987)Uehara, and Horiai]{uehara1987}
Uehara,~H.; Horiai,~K. Infrared-microwave double resonance of atomic chlorine
  on laser-magnetic-resonance fine-structure transitions. \emph{J. Opt. Soc.
  Am. B} \textbf{1987}, \emph{4}, 1217--1221\relax
\mciteBstWouldAddEndPuncttrue
\mciteSetBstMidEndSepPunct{\mcitedefaultmidpunct}
{\mcitedefaultendpunct}{\mcitedefaultseppunct}\relax
\EndOfBibitem
\bibitem[Lekien and Marsden(2005)Lekien, and Marsden]{tricubicspline}
Lekien,~F.; Marsden,~J. Tricubic interpolation in three dimensions.
  \emph{International Journal for Numerical Methods in Engineering}
  \textbf{2005}, \emph{63}, 455--471\relax
\mciteBstWouldAddEndPuncttrue
\mciteSetBstMidEndSepPunct{\mcitedefaultmidpunct}
{\mcitedefaultendpunct}{\mcitedefaultseppunct}\relax
\EndOfBibitem
\bibitem[Duncan \latin{et~al.}(1986)Duncan, Nivellini, and Tullini]{duncan1986}
Duncan,~J.; Nivellini,~G.; Tullini,~F. Methylene chloride: The mid-infrared
  spectrum of an almost vibrationally unperturbed molecule. \emph{Journal of
  Molecular Spectroscopy} \textbf{1986}, \emph{118}, 145--162\relax
\mciteBstWouldAddEndPuncttrue
\mciteSetBstMidEndSepPunct{\mcitedefaultmidpunct}
{\mcitedefaultendpunct}{\mcitedefaultseppunct}\relax
\EndOfBibitem
\bibitem[Liu \latin{et~al.}(2020)Liu, Horton, Yang, Nunes, Shen, Wolf, Forbes,
  Cheng, Moore, Centurion, Hegazy, Li, Lin, Stolow, Hockett, Rozgonyi,
  Marquetand, Wang, and Weinacht]{liu2020}
Liu,~Y. \latin{et~al.}  Spectroscopic and Structural Probing of Excited-State
  Molecular Dynamics with Time-Resolved Photoelectron Spectroscopy and
  Ultrafast Electron Diffraction. \emph{Phys. Rev. X} \textbf{2020}, \emph{10},
  021016\relax
\mciteBstWouldAddEndPuncttrue
\mciteSetBstMidEndSepPunct{\mcitedefaultmidpunct}
{\mcitedefaultendpunct}{\mcitedefaultseppunct}\relax
\EndOfBibitem
\bibitem[Borin \latin{et~al.}(2016)Borin, Matveev, Budkina, El-Khoury, and
  Tarnovsky]{borin2016}
Borin,~V.~A.; Matveev,~S.~M.; Budkina,~D.~S.; El-Khoury,~P.~Z.;
  Tarnovsky,~A.~N. Direct photoisomerization of CH2I2vs. CHBr3 in the gas
  phase: a joint 50 fs experimental and multireference resonance-theoretical
  study. \emph{Phys. Chem. Chem. Phys.} \textbf{2016}, \emph{18},
  28883--28892\relax
\mciteBstWouldAddEndPuncttrue
\mciteSetBstMidEndSepPunct{\mcitedefaultmidpunct}
{\mcitedefaultendpunct}{\mcitedefaultseppunct}\relax
\EndOfBibitem
\bibitem[Dörner \latin{et~al.}(2000)Dörner, Mergel, Jagutzki, Spielberger,
  Ullrich, Moshammer, and Schmidt-Böcking]{dorner2000}
Dörner,~R.; Mergel,~V.; Jagutzki,~O.; Spielberger,~L.; Ullrich,~J.;
  Moshammer,~R.; Schmidt-Böcking,~H. Cold Target Recoil Ion Momentum
  Spectroscopy: a ‘momentum microscope’ to view atomic collision dynamics.
  \emph{Physics Reports} \textbf{2000}, \emph{330}, 95--192\relax
\mciteBstWouldAddEndPuncttrue
\mciteSetBstMidEndSepPunct{\mcitedefaultmidpunct}
{\mcitedefaultendpunct}{\mcitedefaultseppunct}\relax
\EndOfBibitem
\bibitem[Werner and Knowles(1985)Werner, and Knowles]{mcscf1}
Werner,~H.; Knowles,~P.~J. {A second order multiconfiguration SCF procedure
  with optimum convergence}. \emph{J. Chem. Phys.} \textbf{1985}, \emph{82},
  5053--5063\relax
\mciteBstWouldAddEndPuncttrue
\mciteSetBstMidEndSepPunct{\mcitedefaultmidpunct}
{\mcitedefaultendpunct}{\mcitedefaultseppunct}\relax
\EndOfBibitem
\bibitem[Knowles and Werner(1985)Knowles, and Werner]{mcscf2}
Knowles,~P.~J.; Werner,~H.-J. An efficient second-order MC SCF method for long
  configuration expansions. \emph{Chem. Phys. Lett.} \textbf{1985}, \emph{115},
  259--267\relax
\mciteBstWouldAddEndPuncttrue
\mciteSetBstMidEndSepPunct{\mcitedefaultmidpunct}
{\mcitedefaultendpunct}{\mcitedefaultseppunct}\relax
\EndOfBibitem
\bibitem[Krishnan \latin{et~al.}(1980)Krishnan, Binkley, Seeger, and
  Pople]{krishnan1980a}
Krishnan,~R.; Binkley,~J.~S.; Seeger,~R.; Pople,~J.~A. Self-consistent
  molecular orbital methods. XX. A basis set for correlated wave functions.
  \emph{J. Chem. Phys.} \textbf{1980}, \emph{72}, 650--654\relax
\mciteBstWouldAddEndPuncttrue
\mciteSetBstMidEndSepPunct{\mcitedefaultmidpunct}
{\mcitedefaultendpunct}{\mcitedefaultseppunct}\relax
\EndOfBibitem
\bibitem[Corrales \latin{et~al.}(2012)Corrales, Gitzinger, González-Vázquez,
  Loriot, de~Nalda, and Bañares]{corrales2012}
Corrales,~M.~E.; Gitzinger,~G.; González-Vázquez,~J.; Loriot,~V.;
  de~Nalda,~R.; Bañares,~L. Velocity Map Imaging and Theoretical Study of the
  Coulomb Explosion of CH3I under Intense Femtosecond IR Pulses. \emph{J. Phys.
  Chem. A} \textbf{2012}, \emph{116}, 2669--2677\relax
\mciteBstWouldAddEndPuncttrue
\mciteSetBstMidEndSepPunct{\mcitedefaultmidpunct}
{\mcitedefaultendpunct}{\mcitedefaultseppunct}\relax
\EndOfBibitem
\bibitem[Liekhus-Schmaltz \latin{et~al.}(2015)Liekhus-Schmaltz, Tenney, Osipov,
  Sanchez-Gonzalez, Berrah, Boll, Bomme, Bostedt, Bozek, Carron, Coffee, Devin,
  Erk, Ferguson, Field, Foucar, Frasinski, Glownia, G{\"u}hr, Kamalov,
  Krzywinski, Li, Marangos, Martinez, McFarland, Miyabe, Murphy, Natan, Rolles,
  Rudenko, Siano, Simpson, Spector, Swiggers, Walke, Wang, Weber, Bucksbaum,
  and Petrovic]{liekhus2015}
Liekhus-Schmaltz,~C.~E. \latin{et~al.}  Ultrafast isomerization initiated by
  X-ray core ionization. \emph{Nature Communications} \textbf{2015}, \emph{6},
  8199\relax
\mciteBstWouldAddEndPuncttrue
\mciteSetBstMidEndSepPunct{\mcitedefaultmidpunct}
{\mcitedefaultendpunct}{\mcitedefaultseppunct}\relax
\EndOfBibitem
\bibitem[Kaderiya(2021)]{balramthesis}
Kaderiya,~B. Imaging photo-induced dynamics in halomethane molecules with
  coincident ion momentum spectroscopy. Ph.D.\ thesis, Kansas State University,
  2021\relax
\mciteBstWouldAddEndPuncttrue
\mciteSetBstMidEndSepPunct{\mcitedefaultmidpunct}
{\mcitedefaultendpunct}{\mcitedefaultseppunct}\relax
\EndOfBibitem
\end{mcitethebibliography}

\end{document}